# Practical Investigation on the Distinguishability of Longa's Atomic Patterns

Sze Hei Li[1,2], Zoya Dyka[1,2], Alkistis Aikaterini Sigourou[1], Peter Langendoerfer[1,2] and Ievgen Kabin[1]
[1] *IHP – Leibniz Institute for High Performance Microelectronics,* Frankfurt (Oder), Germany
[2] *BTU Cottbus-Senftenberg,* Cottbus, Germany
szehei.li@b-tu.de

*Abstract*—This paper investigates the distinguishability of the atomic patterns for elliptic curve point doubling and addition operations proposed by Longa [2]. We implemented a binary elliptic curve scalar multiplication *kP* algorithm with Longa's atomic patterns for the NIST elliptic curve P-256 using the open-source cryptographic library FLECC in C. We measured and analysed an electromagnetic trace of a single *kP* execution on a microcontroller (TI Launchpad F28379 board). Due to various technical limitations, significant differences in the execution time and the shapes of the atomic blocks could not be determined. Further investigations of the side channel analysis-resistance can be performed based on this work. Last but not least, we examined and corrected Longa's atomic patterns corresponding to formulae proposed by Longa.

## I. Introduction

Cryptographic protocols serve as essential mechanisms for fulfilling the security requirements of communication devices. However, implementing these protocols in embedded devices presents significant challenges due to the resource-constraints of the devices and their susceptibility to physical access by potential attackers. Thus, it is expected that the cryptographic protocols implemented on such devices exhibit minimal execution time, low energy consumption, and robustness against a broad spectrum of physical attacks, particularly horizontal side-channel analysis (SCA) attacks.

SCA attacks, first introduced in 1998 [1], exploit the power consumption patterns of a device during cryptographic operations to reveal the secret cryptographic key. EC-based cryptographic protocols are frequently targeted by attackers aiming to reveal the scalar *k* in the *kP* operation, which refers to the EC scalar multiplication. *kP* operation, being computationally time- and energy- consuming, is typically implemented as a sequence of EC point doubling and point addition operations, which can be further represented as a sequence of mathematical operations with elements of a finite field. Nowadays, there are numerous *kP* algorithms and open-source cryptographic libraries available to designers, facilitating the fast implementation of *kP* algorithms for embedded devices. The SCA resistance of these implementations largely depends on the selected *kP* algorithm and the countermeasures used. The side-channel atomicity principle, introduced by Chevallier-Mames et al. [4], is a well-known countermeasure against horizontal, or single-trace, attacks.

Based on [4], Longa modified the formulae for Elliptic Curve (EC) point doubling and point addition operations and represented them as atomic patterns [2] , which were up to 22% faster than Chevallier-Mames et al.'s atomic patterns. However, practical implementation and analysis of the existing atomic pattern algorithms, especially on embedded devices or as hardware accelerators of EC cryptographic protocols, are rare. This study addresses this gap by implementing Longa's MNAMNAA-based atomic pattern algorithms on a microcontroller (TI Launchpad F28379 board) using the open-source cryptographic library FLECC in C and evaluating the resistance of the implementation to a horizontal side-channel analysis attack.

This paper contributes by:

- Investigating if the distinguishability of Longa's atomic patterns [2] was evaluated experimentally in the past, by reviewing all papers citing [2]
- Correcting Longa's atomic patterns corresponding to Longa's formulae [2]
- Implementing a binary EC scalar multiplication algorithm using Longa's MNAMNAA atomic patterns on a TI Launchpad F28379 evaluation board TMS320F28379D with a 32-bit microcontroller using the open-source cryptographic library FLECC in C
- Investigating the distinguishability of the atomic patterns by analysing the measured electromagnetic trace of an EC point multiplication for the EC P-256 standardized by NIST [5]

The remaining sections of this paper are structured as follows. In Section II, we provide an overview of literature most relevant to the improvement, implementation or analysis of Longa's atomic pattern algorithms [2]. In Section III, we describe the implementation details of the atomic pattern *kP* algorithm using FLECC in C and corrected the names of the registers in Longa's atomic patterns for the EC point doubling and point addition calculations. Section IV describes our experimental setup and measurements. Section V describes the SCA we performed. Finally, Section VI concludes our contributions and limitations.

## II. Overview

Using Google Scholar and Semantic Scholar as search engines, we found and studied 84 scientific papers (see [6], [7]) which cited Longa's work [2]. Only 63 of them are different papers citing [2]. 23 of the 63 papers focus on SCA attacks and/or SCA countermeasures, whereby 15 papers cite [2] only as a state-of-the-art countermeasure against simple SCA attacks; 6 papers claim that Longa's (or others') atomic patterns are possibly vulnerable to horizontal address-bit SCA. Only 10 publications describe improvements, implementations and analysis of Longa's atomic pattern algorithms, or investigate another atomic pattern claiming that their results are applicable to Longa's atomic patterns as well. A short overview of these publications is as follows.

Giraud and Verneuil [8] claimed that Longa's atomic patterns cannot defeat Differential Side-Channel Analysis (DSCA), as the DSCA countermeasures such as projective coordinates randomization [3] or the random curve selection [9] cannot be applied. They improved Longa's algorithms by introducing new fast doubling, general doubling and re-addition algorithms using Longa's atomic patterns. Using these new algorithms for EC point operations, they proposed a right-to-left mixed scalar multiplication algorithm. It was proven to be 10% more efficient than any previous methods at the time. Later, Verneuil published more details regarding the proposed atomic patterns in his PhD thesis [10].

Lu et al. [11] introduced a general framework of side-channel atomicity to support the proposed τ-scalar, ξ-base representation for scalar multiplication, which has shown improved resistance to simple power analysis and supports scalar multiplication algorithms using Longa's atomic patterns. However, Lu et al. only implemented Chevallier-Mames et al.'s atomic pattern "MNAA" [4] due to memory space limitation when implemented in hardware. They claimed that their algorithm hypothetically can accommodate Longa's atomic pattern without changing its effectiveness.

Rondepierre [12] proposed new atomic patterns for EC point doubling and point addition operations and used the patterns for double scalar multiplication using the Straus-Shamir trick [13], [14]. It was shown that the implementation on a smart card has an efficiency improvement of 40% when compared to Giraud and Verneuil's algorithm implementation. He expected that this optimization can also be applied on Longa's pattern. The atomic patterns proposed in [12] can be used for the implementation of a single-scalar $k\bm{P}$ algorithm, too.

Bauer et al. [15] introduced HCCA, and performed a theoretical analysis on Longa's atomic pattern point doubling and point addition algorithms. They pointed out that Longa's algorithms should be susceptible to HCCA. They demonstrated that HCCA against Chevallier-Mames et al.'s atomic pattern algorithm works for 8- and 32-bit microprocessors and ECs of size 160, 250 and 384 bits.

Das et al. [16] recalled the proposition made by Bauer et al. in [15] that Longa's atomic pattern $k\bm{P}$ algorithm is vulnerable to HCCA, since the operand-sharing property among field operations can be detected. They observed that side-channel leakage between two field multiplications can be quantified by computing the distance between two leakages using Pearson correlation coefficient. In addition, they proposed a new protected atomic pattern to demonstrate how to make Giraud and Verneuil's right-to-left atomic scheme safe against HCCA and the improved Big Mac attack [17], with a small overhead.

Kabin et al. [18], [19] implemented in hardware signature generation as well as verification as proposed by Rondepierre [12], using his atomic patterns, and performed a horizontal SCA attack adapting the technique *comparison to the mean* to atomic patterns. Additionally, it was mentioned that all atomic pattern algorithms may be vulnerable to horizontal address-bit SCAs, at least when implemented in hardware.

Kabin [20], Kabin et al. [21] and Sigourou et al. [22] mentioned Longa's atomic patterns as a state-of-the-art countermeasure against simple SCA. Each of these papers investigated the resistance of different $k\bm{P}$ implementations to horizontal (address-bit) SCA attacks. Successful SSCA against an atomic pattern $k\bm{P}$ algorithm is reported in [20] and [22], and against Montgomery ladder in [20]. In [21], it was shown that all algorithmic countermeasures based on randomisation of register addresses for Montgomery ladder [23], [24] and [25] proposed in the past are not effective against horizontal address-bit SCA. Longa's atomic patterns were not implemented or investigated in [20], [21] and [22].

Our literature study shows that none of the papers reported about the implementation of Longa's atomic pattern algorithms in the past. Thus, the indistinguishability of Longa's atomic pattern was never investigated experimentally by analysing measured power or electromagnetic (EM) traces.

### III. IMPLEMENTATION

As the target platform in our experiments, we used a TI LAUNCHXL-F28379D LaunchPad, along with a TMS320F28379D C2000 32-bit dual-core microcontroller as our target device. The microcontroller has 204 kB random-access memory (RAM), 1 MB Flash memory, and operates real-time maximum at 200 MHz. We used Code Composer Studio (CCS) version 12.4.0.00007 as our integrated development environment for the microcontroller.

#### A. Selection of an Open-source Cryptographic Library

A recent work by Sigourou et al. [22] has explored the five most popular open-source cryptographic libraries at the time regarding their suitability for implementation of ECC protocols on resource constrained devices. Their study shows that FLECC in C (FLECC) [26] is the best suited library among others, considering its suitability for embedded devices in terms of source code language used, portability to the attacked device in terms of memory size, access to modular arithmetic functions and presence of constant-runtime functions. Therefore, we selected FLECC for our implementation.

#### B. Atomic Pattern kP Algorithm

We implemented the binary double-and-add left-to-right $k\bm{P}$ algorithm, see Algorithm 1, and used Longa's [2] MNAMNAA-based atomic patterns for EC point doublings and additions.

**Algorithm 1:** Left-to-right binary double-and-add $k\bm{P}$ algorithm
**Input:** $k = (k_{l-1}, ..., k_0)_2, \bm{P} \in E(\mathrm{F}_p)$
**Output:** $k\bm{P}$
1 $Q = P$
2 **for** $i = (l-2)$ *downto* $0$ **do**
3 $\quad Q = 2Q$ // Point doubling atomic pattern
4 $\quad$ **if** $k_i = 1$ **then**
5 $\quad\quad Q = Q + P$ // Point addition atomic pattern
6 **return** $Q$

During the implementation, we found erroneous registers used in Longa's atomic patterns for point doubling and point addition. We examined the correspondence of the atomic patterns to formulae proposed in [2] and corrected the atomic patterns, as shown in Table I and Table II: the corrected registers are marked in bold with yellow highlight. The correctness of all the other atomic patterns proposed in [2] was examined, too. The corrected patterns can be found in [27].

TABLE I.  MNAMNAA ATOMIC PATTERNS SEQUENCE FOR EC POINT DOUBLING IMPLEMENTED IN THIS WORK

Input: $P = (X_1,Y_1,Z_1) \rightarrow (T_1,T_2,T_3)$

|   | Δ1 | Δ2 | Δ3 | Δ4 |
|---|---|---|---|---|
| M | $T_4 \leftarrow T_3^2$ | $T_5 \leftarrow T_4 \cdot T_5$ | $T_5 \leftarrow T_4^2$ | $T_2 \leftarrow T_2^2$ |
| N | * | * | * | $T_5 \leftarrow -T_1$ |
| A | $T_5 \leftarrow T_1 + T_4$ | $T_4 \leftarrow T_5 + T_5$ | $T_6 \leftarrow T_2 + T_2$ | $T_5 \leftarrow T_5 + T_6$ |
| M | $T_6 \leftarrow T_2^2$ | $T_3 \leftarrow T_2 \cdot T_3$ | $T_6 \leftarrow T_1 \cdot T_6$ | $T_5 \leftarrow T_4 \cdot T_5$ |
| N | $T_4 \leftarrow -T_4$ | * | $T_1 \leftarrow -T_6$ | $T_2 \leftarrow -T_2$ |
| A | $T_2 \leftarrow T_2 + T_2$ | $T_4 \leftarrow T_4 + T_5$ | $T_1 \leftarrow T_1 + T_1$ | $T_2 \leftarrow T_2 + T_2$ |
| A | $T_4 \leftarrow T_1 + T_4$ | $T_2 \leftarrow T_6 + T_6$ | $T_1 \leftarrow T_1 + T_5$ | $T_2 \leftarrow T_2 +$ **$T_5$** |

Output: $2P = (X_3,Y_3,Z_3) \leftarrow (T_1,T_2,T_3)$

TABLE II.  MNAMNAA ATOMIC PATTERNS SEQUENCE FOR EC POINT ADDITION IMPLEMENTED IN THIS Work

Input: $P = (X_1,Y_1,Z_1) \rightarrow (T_1,T_2,T_3)$ and $Q = (X_2,Y_2) \rightarrow (T_x,T_y)$

|   | Δ1 | Δ2 | Δ3 |
|---|---|---|---|
| M | $T_4 \leftarrow T_3^2$ | $T_6 \leftarrow T_5^2$ | $T_9 \leftarrow T_5 \cdot T_6$ |
| N | * | * | * |
| A | * | * | $T_8 \leftarrow T_8 + T_9$ |
| M | $T_5 \leftarrow T_x \cdot T_4$ | $T_7 \leftarrow T_1 \cdot T_6$ | $T_4 \leftarrow T_3 \cdot T_4$ |
| N | $T_6 \leftarrow -T_1$ | * | * |
| A | $T_5 \leftarrow T_5 + T_6$ | $T_8 \leftarrow$ **$T_7$** $+$ **$T_7$** | * |
| A | * | * | * |

|   | Δ4 | Δ5 | Δ6 |
|---|---|---|---|
| M | $T_4 \leftarrow T_y \cdot T_4$ | $T_8 \leftarrow T_2 \cdot T_9$ | $T_3 \leftarrow T_3 \cdot T_5$ |
| N | $T_{10} \leftarrow -T_2$ | $T_6 \leftarrow -T_1$ | $T_4 \leftarrow -T_7$ |
| A | $T_4 \leftarrow T_4 + T_{10}$ | $T_6 \leftarrow T_6 + T_7$ | $T_4 \leftarrow T_1 + T_4$ |
| M | $T_{10} \leftarrow T_4 \cdot T_4$ | $T_6 \leftarrow T_6 \cdot$ **$T_4$** | $T_5 \leftarrow T_4 \cdot T_4$ |
| N | $T_8 \leftarrow -T_8$ | $T_9 \leftarrow -T_8$ | $T_6 \leftarrow -T_8$ |
| A | $T_1 \leftarrow$ **$T_{10}$** $+ T_8$ | $T_2 \leftarrow T_6 + T_9$ | $T_6 \leftarrow T_2 + T_6$ |
| A | * | * | * |

Output: $P + Q = (X_3,Y_3,Z_3,X_1',Y_1') \leftarrow (T_1,T_2,T_3,T_4,T_5)$

Each Δ$i$ with $i \in \{1, ..., 6\}$ in Table I and Table II represents an atomic block. Within an atomic block, each line contains one field operation, which is either a multiplication (M), a negation (N) or an addition (A). Each * represents a dummy field operation, making all atomic blocks side-channel indistinguishable. Following the atomic pattern MNAMNAA, point doubling can be computed in 4 atomic blocks, and point addition in 6. We implemented Algorithm 1 applying atomic patterns as shown in Table I and Table II.

The implementation was done for the secp256r1 elliptic curve in FLECC, which is the EC denoted as P-256 by NIST [5]. For calculation of a single field product, we used the constant-time Montgomery modular multiplication function from FLECC twice, thereby we denote the first Montgomery modular multiplication X and the second one X'. All implementation steps including dummy operations are shown in the Appendix.

## IV. MEASUREMENTS

We measured the EM emanation of our implementation during a single $kP$ algorithm execution to provide data for the analysis. For the experiment setup, we used a near-field micro probe (Langer MFA-R 0.2-75 [28]) to measure EM emanation, an integrated circuit scanner (Langer ICS 105 [29]) to precisely position the board and the probe, and an oscilloscope (Teledyne LeCroy WavePro 604HD [30]) to capture the EM emanation. We placed the probe at the side of the capacitor C78 on the attacked board [31] for all of our measurements, as it gave us the strongest EM signals. Fig. 1 shows the attacked board with the microprobe placed.

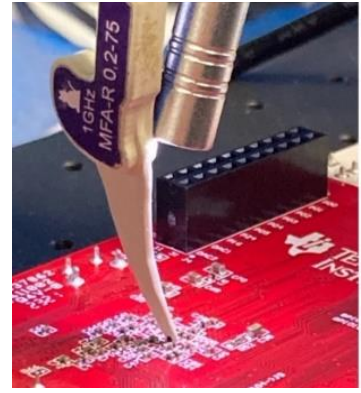

Fig. 1.  The microprobe placed on the attacked board.

For the $kP$ algorithm executed on RAM, we captured the trace with an oscilloscope sampling rate of 1 GS/s. By using a board with 100 MHz clock signal frequency, it results in only 10 samples per clock cycle and a trace file of 6.02 GB. In our experiments, we used the 22-bit long binary scalar $k$= 1001101101011111110111 and the base point $G$ of the EC P-256 [5] as the inputs for our implemented $kP$ algorithm.

In total, 15 point additions and 21 point doublings are performed executing our $kP$ operation with the inputs mentioned above. However, since the full length of the $kP$ operation exceeds the memory limit by the used oscilloscope and its settings, we were not able to capture the last point addition completely[1]. We were only able to capture 14 point additions and 21 point doublings. Furthermore, we used breakpoints in CCS to identify the start of the $kP$ operation executed in RAM as well as for determining the execution time of EC point doublings, additions and the field operations. Table III summarizes our measurement settings and Table IV shows the duration of the first atomic block of each EC point doubling (PD) and point addition (PA) measured in clock cycles using breakpoints.

TABLE III.  SETTINGS APPLIED IN OUR MEASUREMENTS

| Setting | Value |
|---|---|
| Frequency of microcontroller attacked | 100 MHz |
| Sampling rate of the oscilloscope | 1 GS/$s$ |
| Samples per clock cycle | 10 |
| Samples captured | 200 MS |
| $kP$ execution time | >200 $ms$ |
| Size of raw data captured | 6.02 GB |

TABLE IV.  DURATION OF SELECTED OPERATIONS (MEASUERD IN CLOCK CYCLES)

| Measured in | 1st X operation in Δ1 duration (count) | 1st atomic block Δ1 duration (count) |
|---|---|---|
| 21 PDs | 16565(13); 16570(8) | 72768(3); 72773(2); 72778(6); 72783(7); 72788(1); 72793(2) |
| 14 PAs | 16565(13); 16570(1) | 72772(6); 72777(6); 72782(2) |

According to our measurements of the execution time, the duration of Δ1 is similar but not identical in all patterns. The execution time of Δ1 in all PD patterns is between [72768…72793] clock cycles, with a median value of 72778 clock cycles; while that of the PA patterns is between [72772…72782] clock cycles, with a median of 72777 clock cycles. We did not investigate which processes caused the small differences in the range of 10 to 25 clock cycles, but these differences could make the synchronization alignment of the sub-traces not trivial (see subsection VA2).

---

[1] Please note that an increased sampling rate or a longer length of the scalar $k$ will result into significantly increased size of the file storing the $kP$ trace, making the visualization of the trace and the preparation of the trace for the analysis more difficult, even if the technical limitations determined by the oscilloscope used would allow to capture such a trace.

## V. AUTOMATED SIMPLE SCA PERFORMED

### A. Preparation of sub-traces for analysis

#### 1) Separation of sub-traces

To simplify the preparation of the measured EM trace for our analysis, we inserted a sequence of no operations (NOPs) after each atomic block in our *kP* implementation, whereby the sequence of NOPs after each point doubling/addition operation is significantly longer than the sequences after atoms. Fig. 2 shows the beginning of the EM trace we captured, showing the noise before the start of the *kP* operation, the processing of the most significant bit of the scalar *k*, and the shapes of some atomic patterns in the *kP* algorithm.

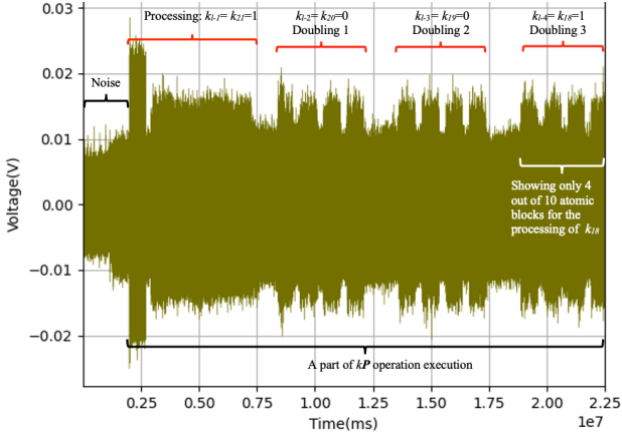

Fig. 2. A part of the captured EM trace.

In the main loop of Algorithm 1, 21 of 22 bits of the scalar *k* are processed, executing 21 point doublings and 15 point additions as atomic patterns. We denote in our analysis all atomic patterns as "Doubling *i*" with $1 \leq i \leq 21$ or "Addition *j*" with $1 \leq j \leq 15$. The numbering starts from the beginning of the *kP* trace. Fig. 3 shows the part of the captured EM trace corresponding to the processing of the bits $k_{l-4} = k_{18} = 1$ and $k_{l-5} = k_{17} = 1$ of the key *k*, with the key length $l=22$.

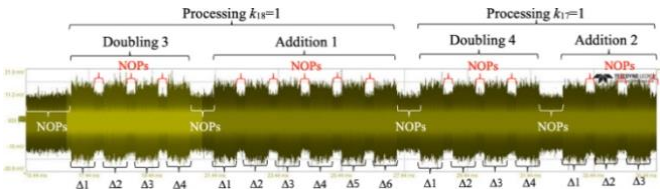

Fig. 3. A part of the measured EM trace corresponding to the processing of $k_{l-4}=k_{18}=1$ and $k_{l-5}=k_{17}=1$.

Processing a key bit value '1' requires the execution of a point doubling and a point addition. Thus, Fig. 3 shows the atomic patterns Doubling 3, Addition 1, Doubling 4 and a part of Addition 2, corresponding to our numbering of the atomic patterns. A point doubling operation contains 4 atomic blocks, whereas a point addition operation contains 6 atomic blocks. The short NOP sequences help us separate the atoms $\Delta 1$ to $\Delta 4$ or $\Delta 1$ to $\Delta 6$. The long NOP sequences help us separate the EC point operations. Additionally, to simplify the separation of the trace into point doubling and point addition sub-traces, we used the information about the duration of operations obtained with breakpoints in the CCS program (see Table IV).

#### 2) Synchronization of sub-traces

We excluded the sub-trace of Doubling 1 from our analysis, due to the special multiplication operand value 1 in its first field multiplication ($Z_1=1 \rightarrow T_3$, before the first atomic block $\Delta 1$, see TABLE I. ). We analysed the first atomic block of 20 doubling sub-traces (Doubling 2 to 21) and 14 addition sub-traces (Addition 1 to 14). All 20+14=34 sub-traces were synchronized manually using Microsoft Excel. Fig. 4 shows the results of synchronization in $\Delta 1$. The red line represents the sub-traces of all point addition operations; the blue line represents that of all point doubling operations; the yellow line marks the mean trace of the point addition sub-traces; the green line marks the mean trace of the point doubling sub-traces; the grey vertical area indicates the anchor samples we used for the fine synchronization.

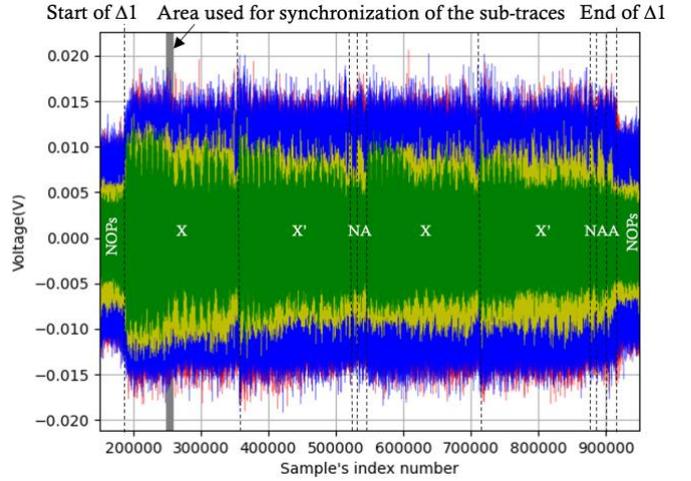

Fig. 4. Overlay of $\Delta 1$ of all sub-traces, after fine-synchronization.

The raw synchronization was done using the information about the start of each $\Delta 1$ obtained by using breakpoints. For the fine synchronization, we selected an interval consisting of 4000 anchor samples in the first half of the first X operation, i.e., the samples with index numbers 252,000…256,000 in Fig. 4. We used the similarities of all shapes among the 200 clock cycles in the anchor interval and aligned local maximums and minimums in all sub-traces. We noticed that the shapes of the red and blue lines are adequately aligned along the anchor interval, see Fig. 5.

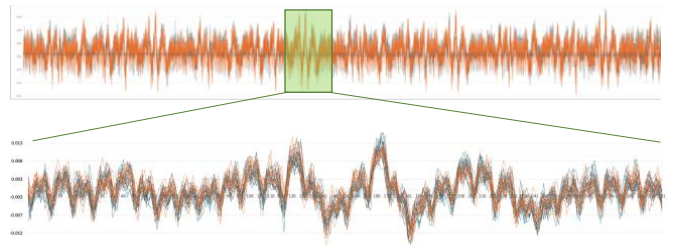

Fig. 5. 4,000 anchor samples (upper part) demonstrating good synchronisation; a zoomed-in part of the anchor interval (lower part).

However, the amplitude of the green line in Fig. 4, i.e., the mean trace of PD sub-traces in $\Delta 1$, diminishes gradually and gets closer to zero as it is further away from the synchronization anchor samples. The yellow line in Fig. 4, i.e., the mean trace of PA sub-traces in Δ1, indicates a better synchronization than that of PD. This can be explained not only by the different number of doubling and addition sub-traces, but also by the different durations of Δ1 for PDs and PAs measured using breakpoints, see Table IV.

*B. Analysis*

We adapted the automated simple SCA method [20] to analyse our synchronized sub-traces. The goal was to determine the samples' index numbers, at which the amplitudes of the set of point addition sub-traces will be completely separated from the set of point doubling sub-traces. The notation "completely separated" means that at sample's index number $i$, the maximum value of all point doubling samples is less than the minimum value of all point addition samples, or the minimum value of all point doubling samples is greater than the maximum value of all point addition samples:

$\max_i$(Doubling 2, …, Doubling 21) < $\min_i$(Addition 1, …, Addition 14)

or

$\max_i$(Addition 1, …, Addition 14) < $\min_i$(Doubling 2, …, Doubling 21)

We developed a Python program to look for the occurrence of such two cases, and examined the sub-traces within the duration of Δ1. Fig. 5*(a)* and *(b)* illustrate our approach. Fig. 5*(b)* shows the maximum-minimum intervals for each atomic pattern during 200 clock cycles within the anchor interval: all blue sub-traces from Fig. 5*(a)* are in the blue area, and all red sub-traces from Fig. 5*(a)* are in the orange area. Our program searches for sample's index, where the orange and blue areas will be completely separated from each other. However, we were unable to find such separations.

Additionally, we used the same program to investigate if the mean traces with their confidence intervals at $\bar{x}\pm\sigma$, $\bar{x}\pm2\sigma$ and $\bar{x}\pm3\sigma$ can be separated from each other, i.e., we performed a kind of difference-of-means test. Fig. 5*(c)* shows the mean traces and their confidence intervals $\bar{x}\pm\sigma$ for the sub-traces from Fig. 5*(a)*. Only few samples were determined as distinguishable samples, for example, the sample with the index number 253,458 pointed at by a red arrow. No separation was observed when comparing both mean traces with their confidence intervals $\bar{x}\pm2\sigma$ or $\bar{x}\pm3\sigma$.

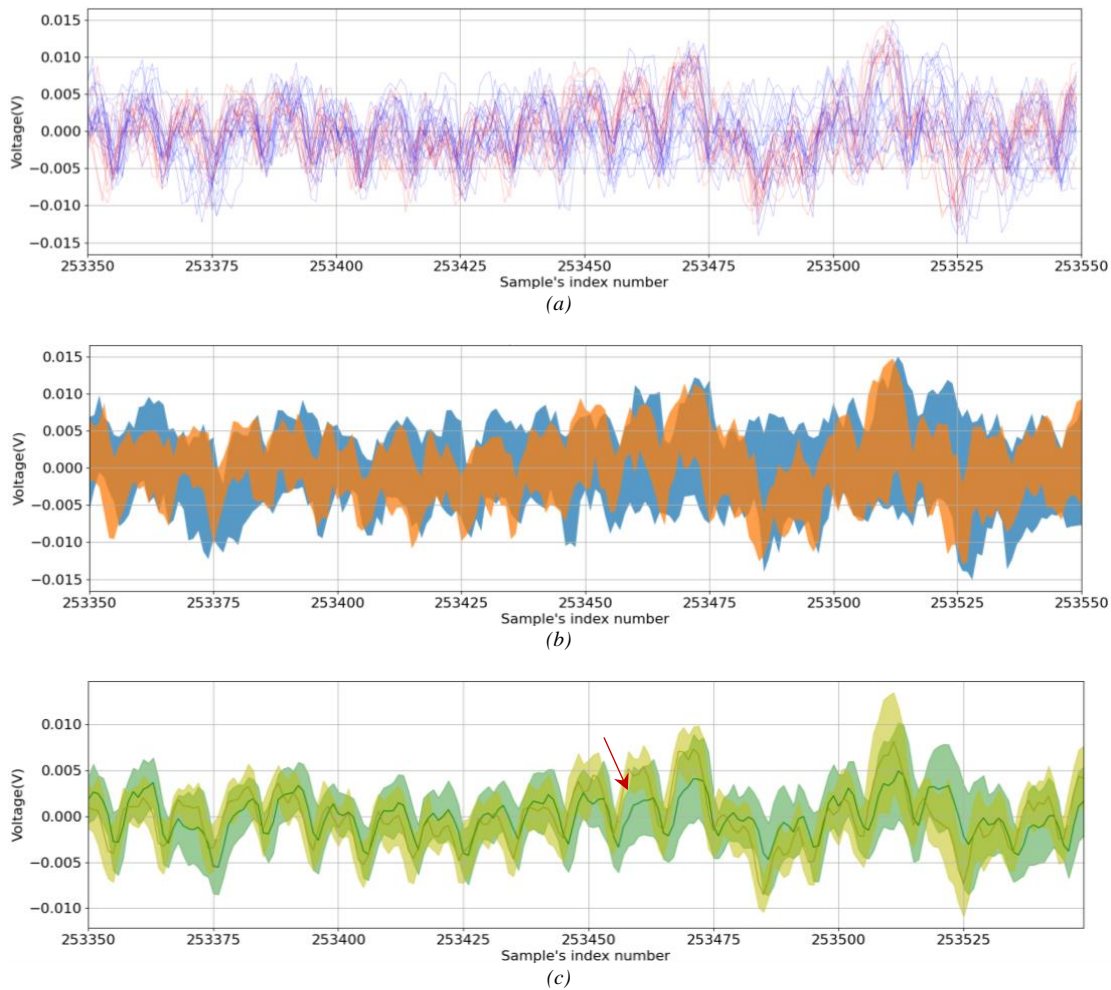

Fig. 5. A part of the synchronized sub-traces within the anchor interval in Δ1: 20 sub-traces of PDs are blue; 14 sub-traces of PAs are red, see *(a)*. The maximum-minimum-areas are shown for the both atomic patterns sets, see *(b)*: all blue sub-traces from *(a)* are in the blue area and all red sub-traces from *(a)* are in the orange area. In *(c)* the both mean traces and their confidence intervals are shown: the bronze line is the mean trace of red sub-traces, and the yellow area shows the corresponding confidence interval $\bar{x}\pm\sigma$; the green line is the mean trace of blue sub-traces, and the green area shows the corresponding confidence interval $\bar{x}\pm\sigma$.

Please note that our expectation was:

- No separation in the anchor interval, due to the fact that this interval refers to a Montgomery modular multiplication, and only the multiplicand values processed are different.
- Separations at the beginning of the 2nd X, 2nd N and 2nd A operation can be a confirmation of the address-bit vulnerability, if observed (see OP4, OP5 and OP6 in Appendix).

*C. Result discussion*

In our investigation, we focused on the analysis of the first atomic block of EC point doubling and point addition atomic patterns. We were unable to determine any clear distinguisher for both operations. The following objective facts made our analysis very difficult:

- The long execution time of the *kP* operation implemented using FLECC library results in a large file size of the measured EM trace:
  - Duration of an atomic block is about 72780 clock cycles.
  - A point doubling consists of 4 atomic blocks, which takes about 377,000 clock cycles; a point addition consists of 6 atomic blocks, which takes more than 580,000 clock cycles.
  By applying these execution time, the minimum execution time of the 256-bit long scalar ($k=k_{255}\ldots k_1 k_0=10\ldots 0$) is 74,190,720 clock cycles, resulting in 742 *ms* by the clock frequency of 100 MHz. The maximum execution time estimated as the time for the processing of $k=k_{255}\ldots k_1 k_0=11\ldots 1$ is 185,476,800 clock cycles, equivalent to almost 2 seconds. Thus, the execution time of a *kP* operation implementing Longa's atomic patterns using FLECC library for the EC P-256 on the investigated microprocessor is within the range from 742 *ms* up to 1855 *ms*. The estimated execution time is very long; this can be a reason why the SCA-resistance of such an implementation was never investigated experimentally.

- Due to the technical limitations, we captured the EM trace only for a 22-bit long scalar with 10 samples per clock cycle. The file containing the trace was 6 GB. About half of the trace are samples corresponding to NOPs, which we inserted to simplify the analysis. But even without the NOPs, a trace of a 256-bit scalar multiplication will be estimated at about 30 GB large.

- High level of noise with small number of samples captured per clock cycle (signal-to-noise ratio of 1.36).

Moreover, our measurements of the execution time of the FLECC constant-time functions used in our implementation show very small differences of execution time, in the range of five clock cycles. But this time difference grows for the atomic block up to 25 clock cycles. These differences in the execution time are not big but they can cause the de-synchronization of the sub-traces.

Due to all these facts, the preparation of the sub-traces for the analysis is an extremely time-consuming process. If the processes which caused the 5 clock cycles long delays could be identified and their shapes could be (automatically) cut out from each sub-trace, the analysis results would be clearer and could answer the question whether Longa's atomic patterns are vulnerable to simple SCA attacks or not. A good synchronisation of all field operation shapes, at least at the beginning of the shapes, is necessary to determine if the addressing of different registers causes the distinguishability of the investigated atomic patterns or not.

The difficulties and limitations listed above have to be considered in future work on the distinguishability of atomic patterns.

## VI. CONCLUSION

This paper presents an investigation of the distinguishability of Longa's atomic patterns [2] when applied to a binary EC *kP* algorithm implementation, assessing its resistance to simple SCA attacks experimentally, i.e., analysing a single measured electromagnetic trace. Our analysis revealed no significant differences in execution time or the shape of the atomic block sub-traces, likely due to technical limitations as well as the difficult and time-consuming synchronization process of the sub-traces. Additionally, this paper corrected the names of the registers in Longa's atomic patterns for the calculations of EC point doubling and point addition. Based on the knowledge of the limitations described in this paper, further investigation is required to conclusively determine whether Longa's atomic patterns in the *kP* algorithm are vulnerable to simple SCA attacks or not.

## APPENDIX

The sequence of operations in our implementation of the atomic patterns including dummy operations:

| Atoms | Operations | EC Point Doubling<br>Input: $P = (X_1, Y_1, Z_1)$<br>Output: $2P = (X_3, Y_3, Z_3)$<br>$T_1 \leftarrow X_1, T_2 \leftarrow Y_1, T_3 \leftarrow Z_1$ | | EC Point Addition<br>Input: $P = (X_1, Y_1, Z_1), Q = (X_2, Y_2)$<br>Output: $P + Q = (X_3, Y_3, Z_3, X_1', Y_1')$<br>$T_1 \leftarrow X_1, T_2 \leftarrow Y_1, T_3 \leftarrow Z_1$,<br>$T_x \leftarrow X_2, T_y \leftarrow Y_2$ | |
|---|---|---|---|---|---|
| | | Atomic pattern corresponding to Table I | Our implementation using constant-time FLECC functions | Atomic pattern corresponding to Table II | Our implementation using constant-time FLECC functions |
| Δ1 | OP1:M | $T_4 \leftarrow T_3 \cdot T_3$ | $T_4 \leftarrow T_3 \cdot T_3 \cdot R^{-1}$<br>$T_4 \leftarrow T_4 \cdot R^2 \cdot R^{-1}$ | $T_4 \leftarrow T_3 \cdot T_3$ | $T_4 \leftarrow T_3 \cdot T_3 \cdot R^{-1}$<br>$T_4 \leftarrow T_4 \cdot R^2 \cdot R^{-1}$ |
| | OP2:N | * | $T_0 \leftarrow -T_1$ | * | $T_0 \leftarrow -T_1$ |
| | OP3:A | $T_5 \leftarrow T_1 + T_4$ | $T_5 \leftarrow T_1 + T_4$ | * | $T_5 \leftarrow T_1 + T_4$ |
| | OP4:M | $T_6 \leftarrow T_2 \cdot T_2$ | $T_6 \leftarrow T_2 \cdot T_2 \cdot R^{-1}$<br>$T_6 \leftarrow T_6 \cdot R^2 \cdot R^{-1}$ | $T_5 \leftarrow T_x \cdot T_4$ | $T_5 \leftarrow T_x \cdot T_4 \cdot R^{-1}$<br>$T_5 \leftarrow T_5 \cdot R^2 \cdot R^{-1}$ |
| | OP5:N | $T_4 \leftarrow -T_4$ | $T_4 \leftarrow -T_4$ | $T_6 \leftarrow -T_1$ | $T_6 \leftarrow -T_1$ |
| | OP6:A | $T_2 \leftarrow T_2 + T_2$ | $T_2 \leftarrow T_2 + T_2$ | $T_5 \leftarrow T_5 + T_6$ | $T_5 \leftarrow T_5 + T_6$ |
| | OP7:A | $T_4 \leftarrow T_1 + T_4$ | $T_4 \leftarrow T_1 + T_4$ | * | $T_0 \leftarrow T_1 + T_4$ |
| Δ2 | OP8:M | $T_5 \leftarrow T_4 \cdot T_5$ | $T_5 \leftarrow T_4 \cdot T_5 \cdot R^{-1}$<br>$T_5 \leftarrow T_5 \cdot R^2 \cdot R^{-1}$ | $T_6 \leftarrow T_5 \cdot T_5$ | $T_6 \leftarrow T_5 \cdot T_5 \cdot R^{-1}$<br>$T_6 \leftarrow T_6 \cdot R^2 \cdot R^{-1}$ |
| | OP9:N | * | $T_0 \leftarrow -T_2$ | * | $T_0 \leftarrow -T_2$ |
| | OP10:A | $T_4 \leftarrow T_5 + T_5$ | $T_4 \leftarrow T_5 + T_5$ | * | $T_0 \leftarrow T_5 + T_5$ |
| | OP11:M | $T_3 \leftarrow T_2 \cdot T_3$ | $T_3 \leftarrow T_2 \cdot T_3 \cdot R^{-1}$<br>$T_3 \leftarrow T_3 \cdot R^2 \cdot R^{-1}$ | $T_7 \leftarrow T_1 \cdot T_6$ | $T_7 \leftarrow T_1 \cdot T_6 \cdot R^{-1}$<br>$T_7 \leftarrow T_7 \cdot R^2 \cdot R^{-1}$ |
| | OP12:N | * | $T_0 \leftarrow -T_4$ | * | $T_0 \leftarrow -T_4$ |
| | OP13:A | $T_4 \leftarrow T_4 + T_5$ | $T_4 \leftarrow T_4 + T_5$ | $T_8 \leftarrow T_7 + T_7$ | $T_8 \leftarrow T_7 + T_7$ |
| | OP14:A | $T_2 \leftarrow T_6 + T_6$ | $T_2 \leftarrow T_6 + T_6$ | * | $T_0 \leftarrow T_0 + T_0$ |
| Δ3 | OP15:M | $T_5 \leftarrow T_4 \cdot T_4$ | $T_5 \leftarrow T_4 \cdot T_4 \cdot R^{-1}$<br>$T_5 \leftarrow T_5 \cdot R^2 \cdot R^{-1}$ | $T_9 \leftarrow T_5 \cdot T_6$ | $T_9 \leftarrow T_5 \cdot T_6 \cdot R^{-1}$<br>$T_9 \leftarrow T_9 \cdot R^2 \cdot R^{-1}$ |
| | OP16:N | * | $T_0 \leftarrow -T_1$ | * | $T_0 \leftarrow -T_1$ |
| | OP17:A | $T_6 \leftarrow T_2 + T_2$ | $T_6 \leftarrow T_2 + T_2$ | $T_8 \leftarrow T_8 + T_9$ | $T_8 \leftarrow T_8 + T_9$ |
| | OP18:M | $T_6 \leftarrow T_1 \cdot T_6$ | $T_6 \leftarrow T_1 \cdot T_6 \cdot R^{-1}$<br>$T_6 \leftarrow T_6 \cdot R^2 \cdot R^{-1}$ | $T_4 \leftarrow T_3 \cdot T_4$ | $T_4 \leftarrow T_3 \cdot T_4 \cdot R^{-1}$<br>$T_4 \leftarrow T_4 \cdot R^2 \cdot R^{-1}$ |
| | OP19:N | $T_1 \leftarrow -T_6$ | $T_1 \leftarrow -T_6$ | * | $T_0 \leftarrow -T_6$ |
| | OP20:A | $T_1 \leftarrow T_1 + T_1$ | $T_1 \leftarrow T_1 + T_1$ | * | $T_0 \leftarrow T_1 + T_1$ |
| | OP21:A | $T_1 \leftarrow T_1 + T_5$ | $T_1 \leftarrow T_1 + T_5$ | * | $T_0 \leftarrow T_1 + T_5$ |
| Δ4 | OP22:M | $T_2 \leftarrow T_2 \cdot T_2$ | $T_2 \leftarrow T_2 \cdot T_2 \cdot R^{-1}$<br>$T_2 \leftarrow T_2 \cdot R^2 \cdot R^{-1}$ | $T_4 \leftarrow T_y \cdot T_4$ | $T_4 \leftarrow T_y \cdot T_4 \cdot R^{-1}$<br>$T_4 \leftarrow T_4 \cdot R^2 \cdot R^{-1}$ |
| | OP23:N | $T_5 \leftarrow -T_1$ | $T_5 \leftarrow -T_1$ | $T_{10} \leftarrow -T_2$ | $T_{10} \leftarrow -T_2$ |
| | OP24:A | $T_5 \leftarrow T_5 + T_6$ | $T_5 \leftarrow T_5 + T_6$ | $T_4 \leftarrow T_4 + T_{10}$ | $T_4 \leftarrow T_4 + T_{10}$ |
| | OP25:M | $T_5 \leftarrow T_4 \cdot T_5$ | $T_5 \leftarrow T_4 \cdot T_5 \cdot R^{-1}$<br>$T_5 \leftarrow T_5 \cdot R^2 \cdot R^{-1}$ | $T_{10} \leftarrow T_4 \cdot T_4$ | $T_{10} \leftarrow T_4 \cdot T_4 \cdot R^{-1}$<br>$T_{10} \leftarrow T_{10} \cdot R^2 \cdot R^{-1}$ |
| | OP26:N | $T_2 \leftarrow -T_2$ | $T_2 \leftarrow -T_2$ | $T_8 \leftarrow -T_8$ | $T_8 \leftarrow -T_8$ |
| | OP27:A | $T_2 \leftarrow T_2 + T_2$ | $T_2 \leftarrow T_2 + T_2$ | $T_{10} \leftarrow T_{10} + T_8$ | $T_1 \leftarrow T_{10} + T_8$ |
| | OP28:A | $T_2 \leftarrow T_2 + T_5$ | $T_2 \leftarrow T_2 + T_5$ | * | $T_0 \leftarrow T_2 + T_5$ |
| Δ5 | OP29:M | | | $T_8 \leftarrow T_2 \cdot T_9$ | $T_8 \leftarrow T_2 \cdot T_9 \cdot R^{-1}$<br>$T_8 \leftarrow T_8 \cdot R^2 \cdot R^{-1}$ |
| | OP30:N | | | $T_6 \leftarrow -T_1$ | $T_6 \leftarrow -T_1$ |
| | OP31:A | | | $T_6 \leftarrow T_6 + T_7$ | $T_6 \leftarrow T_6 + T_7$ |
| | OP32:M | | | $T_6 \leftarrow T_6 \cdot T_4$ | $T_6 \leftarrow T_6 \cdot T_4 \cdot R^{-1}$<br>$T_6 \leftarrow T_6 \cdot R^2 \cdot R^{-1}$ |
| | OP33:N | | | $T_9 \leftarrow -T_8$ | $T_9 \leftarrow -T_8$ |
| | OP34:A | | | $T_2 \leftarrow T_6 + T_9$ | $T_2 \leftarrow T_6 + T_9$ |
| | OP35:A | | | * | $T_0 \leftarrow T_2 + T_5$ |
| Δ6 | OP36:M | | | $T_3 \leftarrow T_3 \cdot T_5$ | $T_3 \leftarrow T_3 \cdot T_5 \cdot R^{-1}$<br>$T_3 \leftarrow T_3 \cdot R^2 \cdot R^{-1}$ |
| | OP37:N | | | $T_4 \leftarrow -T_7$ | $T_4 \leftarrow -T_7$ |
| | OP38:A | | | $T_4 \leftarrow T_1 + T_4$ | $T_4 \leftarrow T_1 + T_4$ |
| | OP39:M | | | $T_5 \leftarrow T_4 \cdot T_4$ | $T_5 \leftarrow T_4 \cdot T_4 \cdot R^{-1}$<br>$T_5 \leftarrow T_5 \cdot R^2 \cdot R^{-1}$ |
| | OP40:N | | | $T_6 \leftarrow -T_8$ | $T_6 \leftarrow -T_8$ |
| | OP41:A | | | $T_6 \leftarrow T_2 + T_6$ | $T_6 \leftarrow T_2 + T_6$ |
| | OP42:A | | | * | $T_0 \leftarrow T_2 + T_5$ |